\documentclass[12pt]{article}
\usepackage{graphics}
\usepackage{array}
\usepackage{graphicx}
\usepackage{amssymb}
\usepackage{amsmath}
\input{epsf}
\usepackage{cite}
\def\@fmsl@sh#1#2#3{\m@th\ooalign{$\hfil#1\mkern#2/\hfil$\crcr$#1#3$}}
 \def\eq#1\en{\begin{equation}#1\end{equation}}
\def\s[#1,#2]{[#1\stackrel{\star}{,}#2]}
\def\sx[#1,#2]{[#1\stackrel{\star_{x}}{,}#2]}

\newcommand{\nc}{\newcommand}
\nc{\beq}{\begin{equation}}
\nc{\eeq}{\end{equation}}
\nc{\beqa}{\begin{eqnarray}}
\nc{\eeqa}{\end{eqnarray}}

\def\bc{\begin{center}}
\def\ec{\end{center}}

\def\gsim{\mathrel{\rlap{\lower4pt\hbox{\hskip1pt$\sim$}}
    \raise1pt\hbox{$>$}}}       %greater than or approx. symbol

\begin{document}
\makeatletter
\def\fmslash{\@ifnextchar[{\fmsl@sh}{\fmsl@sh[0mu]}}
\def\fmsl@sh[#1]#2{%
  \mathchoice
    {\@fmsl@sh\displaystyle{#1}{#2}}%
    {\@fmsl@sh\textstyle{#1}{#2}}%
    {\@fmsl@sh\scriptstyle{#1}{#2}}%
    {\@fmsl@sh\scriptscriptstyle{#1}{#2}}}
\def\@fmsl@sh#1#2#3{\m@th\ooalign{$\hfil#1\mkern#2/\hfil$\crcr$#1#3$}}
\makeatother
%\baselineskip 24pt

%%%%%%%%%%%%%%%%%%%%%%%%%%%%%%%%%%%%%%%%%%%%%%%%%%%%%%%%%%%%%%%%%
%%%
%%%                      TITLE PAGE
%%%
%%%%%%%%%%%%%%%%%%%%%%%%%%%%%%%%%%%%%%%%%%%%%%%%%%%%%%%%%%%%%%%%%

\title{Quantum Gravity at the LHC}  

\author{Xavier Calmet\thanks{xcalmet@gmail.com} $^a$ \ and
Priscila de Aquino\thanks{priscila@itf.fys.kuleuven.be} $^{b,c}$ 
 \\
$^a$  Physics and Astronomy, 
University of Sussex \\ Falmer, Brighton,  BN1 9QH, United Kingdom \\
 $^b$ Universit\' e catholique de Louvain \\
Center for Particle Physics and Phenomenology \\
2, Chemin du Cyclotron\\
B-1348 Louvain-la-Neuve, Belgium\\ 
$^c$ Katholieke Universiteit Leuven,\\
Institute of Theoretical Physics\\
Celestijnenlaan 200D - Bus 2415\\
B-3001 Leuven, Belgium 
}

\date{September 2009}

\maketitle

\begin{abstract}
It has recently been shown that if there is a large hidden sector in Nature, the scale of quantum gravity could be much lower than traditionally expected. We study the production of massless  gravitons at the LHC and compare our results to those obtained in extra dimensional models. The signature in both cases is missing energy plus jets.  In case of non observation, the LHC could be used to put the tightest limit to date on the value of the Planck mass.
\end{abstract}

%\pacs{}

%%%%%%%%%%%%%%%%%%%%%%%%%%%%%%%%%%%%%%%%%%%%%%%%%%%%%%%%%%%%%%%%
%%%
%%%                     INTRODUCTION
%%%
%%%%%%%%%%%%%%%%%%%%%%%%%%%%%%%%%%%%%%%%%%%%%%%%%%%%%%%%%%%%%%%%

\newpage
\section{Introduction}

The Planck mass  which is defined as $M=\sqrt{\hbar c/G_N} \approx 1.2209 \times 10^{19} \mbox{GeV}/c^2$, where $\hbar$ is Planck's constant, $c$ is the speed of light and $G_N$ is Newton's constant, is typically assumed to be the energy scale at which quantum gravitational effects become important\footnote{In the sequel, we shall set $\hbar=c=1$.}. However, this definition for the energy scale of quantum gravity could be too naive. Indeed, it has been shown that if there are more than four spacetime dimensions in Nature, for example in models with brane worlds and a large extra dimensional volume, the true scale at which gravity becomes strong could be much lower than naively assumed \cite{ArkaniHamed:1998rs, Randall:1999ee}.  Even in four dimensions, the Planck mass could be much lower than $10^{19}$ GeV. It has recently been realized that the renormalization of the Newton's constant and hence of the Planck mass could lead to strong gravitational effects in the TeV region in four spacetime dimensions if there is a large hidden sector that interacts only gravitationally with the Standard Model of particle physics \cite{Calmet:2008tn}.

Einstein's dream of an unification of all forces of Nature including gravity is still very far away, however much progress has been done in understanding how to formulate quantum field theories in curved spacetime \cite{Birrell:1982ix} and in treating general relativity as an effective field theory (see e.g. \cite{Donoghue:1994dn}). We are used to think of the Planck scale $M$ as a fundamental scale of Nature in which quantum gravitational effects become important. However, this coupling constant gets renormalized when quantum fluctuations are taken into account like any other coupling constant or mass parameter of a quantum field theory. In other words, Newton's constant and hence the Planck mass  are scale dependent. Then  the true scale $\mu_*$ at which quantum gravity effects are large is one at which
\begin{equation}
\label{strong}
M (\mu_*) \sim \mu_*.
\end{equation} 
This condition implies that  quantum fluctuations in spacetime geometry at length scales $\mu_*^{-1}$ will be unsuppressed. One can think of $\mu_*^{-1}$  as the minimal measurable length in Nature\cite{Calmet:2004mp,Calmet:2007vb}.

In this paper,  we shall first  describe  how a large hidden sector with some $10^{33}$ particles of spin 0 and spin 1/2 can lead to a running of Newton's constant and to a scale of quantum gravity $\mu_*$ in the TeV region. The aim of this work is to consider the phenomenology of this model at the LHC. It has been shown that AGASA, a cosmic rays experiment, implies a bound of roughly 550 GeV on the Planck mass in four dimensions \cite{Calmet:2008rv}. The derivation of this bound assumes that neutrinos are the dominant component of high energy cosmic rays. If this turns out not to be the case, then this bound is void. In any case, the important information is that there is no tight bound on the value of the Planck mass in four dimensions and it could be relevant for LHC physics assuming gravity is involved in the solution to the hierarchy problem of the Standard Model.  One of the consequences of this model is that small, quantum black holes might be produced at the LHC. This has been considered in details in  \cite{Calmet:2008dg}. In the present paper we shall consider the massless graviton emission in proton proton collision at the LHC and compare our four dimensional model to the large extra dimensional one. We point out that the LHC will allow either to discover a large hidden sector that interacts only gravitationally with the Standard Model or to put the tightest experimental bound to date on the scale at which quantum gravity becomes relevant. This bound would be more reliable than the one obtained using quantum black holes since the latter one will depend on a series of assumptions concerning quantum black holes.

This paper is organized as follow. We first describe the four dimensional model with a large hidden sector and explain how this sector can impact the running of the Planck mass. We then calculate the parton level cross sections necessary to describe the reaction proton+proton $\to$ graviton + jets, where the graviton is massless and which is relevant for the LHC, i.e. $\bar q + q \to G+ g$,  $q+g \to G+ q$,  $\bar q + g \to G + \bar q$ and  $g+g \to G + g$. We compare our results to those obtained for the emission of massive Kaluza Klein gravitons at the LHC \cite{Giudice:1998ck,Han:1998sg,Li:2006yv} . Generically speaking, it is possible to obtain the parton level cross sections for the massless graviton from the massive graviton case by taking the mass of the Kaluza Klein graviton to zero. Finally,  we  conclude.

\section{A large hidden sector}

\begin{figure}
%[htp]
\centering
\includegraphics[ width=3in]{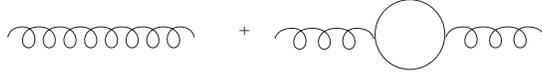}
\caption{Contributions to the running of Newton's constant}\label{Figure2}
\end{figure}

As mentioned in the introduction, Newton's constant and hence the Planck mass do get renormalized by virtual particles. Consider the gravitational potential between two heavy, non relativistic sources, which arises through graviton exchange (see Figure (\ref{Figure2})).  The virtual particles will renormalize the coupling of graviton to the heavy degrees of freedom and hence Newton's constant.
 The actual calculations are done using the heat kernel technique \cite{Larsen:1995ax,Kabat:1995eq,DeWitt:1965jb} which is not based on Feynman diagrams.

Technically speaking, a Wilsonian Planck mass $M( \mu )$ can be introduced. The contributions of spin 0, spin 1/2 and spin 1 particles  to the running of $M( \mu )$  can be calculated using the heat kernel method. This regularization procedure insures that the symmetries of the theory are preserved by the regulator. One finds  \cite{Calmet:2008tn} 
 \begin{eqnarray}
 \label{Nrunning}
M ( \mu )^2=M(0)^2 - \frac{\mu^2}{12 \pi} N
\end{eqnarray}
where $M(0)$ is the Planck mass measured in long distance (astrophysical) experiments and where $N=N_0+N_{1/2}-4 N_1$. The parameters $N_0$,  $N_{1/2}$ and  $N_1$ are respectively  the number of scalars, the number of  Weyl fermions and the number of gauge bosons in the theory. Note that this calculation relies on quantum field theory in curved spacetime and does not require any assumption about quantum gravity. Furthermore, as noted in \cite{Kabat:1995eq}, the contribution of the photon is gauge independent. 

An unification of all forces of Nature is particularly difficult to be realized because of different factors. A technical one is the question of the exact formulation of a theory of quantum gravity. A more physical one may be the apparent weakness of gravity compared to the gauge interactions of the Standard Model. A solution to that problem could simply be that gravity becomes comparable in strength to the other fundamental interactions of Nature because of its running. A natural scale to expect the unification could be somewhere between the $10^{16}$ GeV and $10^{19}$ GeV, which is between the grand unification scale and the traditional Planck mass, but it could also be much lower if quantum gravity is linked to a solution of the hierarchy problem of the Standard Model. In that case, gravity has to become comparable in strength to the strong, weak and electromagnetic interactions around 1 TeV.

Using equations  (\ref{Nrunning}) and (\ref{strong}), one finds that $\mu_\star \sim 1$ TeV requires $N=5.6 \times 10^{33}$ new particles of spin 0 or spin 1/2 or a combination of both. As explained in \cite{Calmet:2008tn}, if this hidden sector is supersymmetric, the cosmological problem does not get worse since the contribution of the hidden sector to the cosmological constant cancels out exactly.
One may worry of the reliability of the calculation when the sliding scale $\mu$ gets close to $\mu_\star$, however we shall argue in appendix A,  that in the large $N$ limit the calculation is under control. In particular, loops involving virtual gravitons are suppressed by $1/N$.

The coupling of the graviton to the Standard Model is the usual one with the understanding that the reduced Planck mass $\bar M$ is now a scale dependent parameter. The linearized theory is then given by
 \begin{eqnarray}
 \label{action}
{\cal L}&=&\frac{1}{4}(\partial^\mu h^{\nu\rho}(x) \partial_\mu h_{\nu\rho}(x) - \partial^\mu h(x) \partial_\mu h(x) - 2 h^\mu(x) h_\mu(x) + 2 h^\mu(x) \partial_\mu h(x)) 
\\ \nonumber &&
-\frac{1}{\bar M(\mu)} h_{\mu\nu}(x) T^{\mu\nu}_{SM}(x)
\end{eqnarray}
where $h_{\mu\nu}(x)$ is the massless graviton, $h(x)=h^\mu_{\ \mu}(x)$, $h_\nu(x)=\partial^\mu h_{\mu\nu}(x)$ and $T^{\mu\nu}_{SM}(x)$ is the energy-momentum tensor of the Standard Model. The Feynman rules can be found in many papers. We follow the convention of  \cite{Han:1998sg} and reproduce the necessary results in appendix B.

\section{Graviton emission at the LHC}

The production of jets with large $E_T$ recoiling against a graviton G can 
arise from the  parton subprocesses  $q+ \bar q \to G + g$, $q+ g \to q+ G$,
 $\bar q + g \to \bar q + G$, and $g+g \to g+ G$. Using the Feynman rules given in appendix B for linearized four-dimensional general relativity coupled to the Standard Model, we have calculated the leading order contributions at the parton level. We have treated all quarks as being massless.
 The polarization and color averaged cross section for $q + \bar q \to g +G$ is given by
\begin{eqnarray}
 \frac{d\sigma}{d \cos\theta}&=& \frac{1}{72 \pi}
 \frac{g_s^2 }{\bar M(\mu)^2} \left ( \frac{s^2 +2t^2+ 2 t s}{s^2} \right ),
 \end{eqnarray}
where $g_s$ is the strong coupling constant, $\bar M(\mu)$ is the reduced Planck Mass and where $s$ and $t$ are the Mandelstam variables: $t= -1/2 s (1 - \cos \theta)$.
The  cross sections for $q + g \to q + G$ and $\bar q + g \to \bar q + G$  are given by
 \begin{eqnarray}
\frac{d \sigma}{d \cos \theta} = - \frac{g_s^2 (2 s^2+ 2 s t + t^2)}{ 192 \pi \bar M(\mu)^2 s t }.
\end{eqnarray}
The corresponding matrix element can be obtained using crossing symmetry from that of the transition $q +\bar q \to G + g$. These cross-sections can be deduced from \cite{Holstein:2006bh}.
Finally we also obtain the cross section for  $g+ g \to g + G$ which is given by
 \begin{eqnarray}
\frac{d \sigma}{d \cos \theta} = - \frac{3 g_s^2 (s^2+s t + t^2)^2}{ 128 \pi \bar M(\mu)^2 s^2 t (s+t)}.
\end{eqnarray}
Note that the non Abelian nature of the QCD interaction is important for this cross section and one cannot sum naively over the polarization of the gluons. One needs to either introduce Fadeev Popov ghosts or restrict the sum to the transverse polarizations, which is the option we follow. We use the standard trick
  \begin{eqnarray}
 \sum_T \epsilon_T^\mu(k)  \epsilon_T^\nu(k)= -\eta^{\mu\nu}+\frac{n^\mu k^\nu+k^\mu n^\nu}{n.k}-\frac{n^2 k^\mu k^\nu}{(n.k)^2}
  \end{eqnarray}
 where $n$ is an arbitrary vector. Note that the calculations can be simplified tremendously by noting that the graviton is on-shell and thus any contraction  of the style $T^{\mu\nu} k_\mu$
 where $T^{\mu\nu}$ is the energy momentum tensor and $k_\mu$ is the momentum of the graviton vanishes.
 The sum over the polarizations of the graviton is given by \cite{vanDam:1970vg}
  \begin{eqnarray}
 \sum _{i=1}^2\epsilon^i_{\mu\nu}(k)  \epsilon^i_{\alpha\beta}(k)= \frac{1}{2}(\eta_{\mu\alpha} \eta_{\nu\beta}+\eta_{\mu\beta} \eta_{\nu\alpha} - \eta_{\mu\nu} \eta_{\alpha\beta}). 
   \end{eqnarray}
 In the sequel we identify the energy scale $\mu$ with the partonic center of mass energy $\sqrt{\hat s}$. For collisions with $\sqrt{\hat s} < \mu_* \sim 1$ TeV, quantum gravity contributions are so weak ($\bar M(\mu) \sim 10^{18}$  GeV for $\sqrt{\hat s} < \mu_*$) that the cross sections go to zero fast. However for $\sqrt{\hat s} > \mu_*$, $\bar M(\mu)\sim$1 TeV and gravitons will be produced. 

We shall implement the running of the Planck mass with a Heaviside step function in the cross section, i.e. the Planck mass for collisions at the parton level with $\sqrt{\hat s}>1$ TeV is given by $\bar M(\mu_*)=\bar \mu_*= 1$ TeV,  but for less energetic parton level collisions  we take $\bar M \to \infty$. The cross section for proton+proton $\to$ Graviton + jets at a center of mass of 14 TeV is  $5 \times 10^5$ fb. For the numerical evaluation we used the mathematica package for  parton distribution functions of the CTEQ collaboration. Obviously the graviton is not detectable and appears as missing energy and  the signature for the emission of  a graviton is then proton + proton $\to$ jets + missing energy.  These signals, although potentially difficult to observe if the background is poorly understood, would be appearing before the quantum black holes discussed in \cite{Calmet:2008dg} and would thus be the first signals of quantum gravity at the LHC. We shall now compare our results with those obtained in extra-dimensional models.

 \section{Comparison with large extra-dimensions models}
 
 At the fundamental level the four dimensional model discussed above and the extra-dimensional models are rather different. However, their phenomenology could be quite similar. First of all, the reason why  an observable effect is expected in the extra-dimensional scenario is that many Kaluza-Klein excitations of the graviton would be produced, however, each individual Kaluza-Klein copy of the graviton couples only with the usual reduced Planck mass (i.e. $10^{18}$ GeV) to the Standard Model particles. The sum runs over some $10^{32}$ Kaluza-Klein states. The sum of the individual partonic cross-sections is thus sizable. This is in sharp contrast to the four dimensional model discussed above. It is interesting to compare the partonic cross sections for the production of a massive graviton to our calculations. Doing a literature survey, we noticed many different and incompatible results for the massive graviton case.  We thus have redone this calculation and we do agree with the work of  Mirabelli {\it et al.} \cite{Mirabelli:1998rt}. We find that the cross section $q + \bar q \to g + G_{KK}$ where $G_{KK}$ is a Kaluza-Klein graviton is given by:
\begin{eqnarray}
\frac{d\sigma}{d \cos\theta}(q + \bar q \to g + G_{KK})  &=& \frac{1}{144 \pi}
 \frac{g_s^2 }{\bar M^2}\frac{1}{1-m^2/s}\Biggl[(2 - \frac{4ut}{(s-m^2)^2}) 
  \left(1 + \bigl(\frac{m^2}{s}\bigr)^4\right) \\ \nonumber
     &&  + \left(2 \frac{ (s-m^2)^2}{4 ut} - 5 
      + 4 \frac{4ut}{(s-m^2)^2}\right)\frac{m^2}{s } \left(1 +  
   \bigl(\frac{m^2}{s}\bigr)^2\right) + 
   \\ \nonumber
     && + 6 \left( \frac{u-t}{s-m^2}\right)^2
                          \bigl(\frac{m^2}{s}\bigr)^2
                                          \Biggr] \ ,
\end{eqnarray}
where $s,t,u$ are the Mandelstam variables we the usual definitions: $t,u = -1/2 s (1-m^2/s)(1\mp 
\cos \theta)$.
The cross section for $q+g \to q+ G_{KK}$ 
 can be obtained from this expression 
by crossing $s \leftrightarrow t$:
\begin{eqnarray}
\frac{d\sigma}{d \cos\theta}(q+g \to q+ G_{KK})  &=&  
\frac {g_s^2}{384 \pi \bar M^2} \frac{(-t/s)(1-m^2/s)}{(1-m^2/t)^2} \times \\
\nonumber
     && 
 \times \Biggl[(2 - \frac{4us}{ (t-m^2)^2}) 
  \left(1 + \bigl(\frac{m^2}{t}\bigr)^4\right) \\ \nonumber
     && + \left(2 \frac{ (t-m^2)^2}{4 us} - 5 
      + 4 \frac{4us}{(t-m^2)^2}\right)\frac{m^2}{t }\left(1 +  
   \bigl(\frac{m^2}{t}\bigr)^2\right) +\\ \nonumber
     && + 6 \left( \frac{s-u}{t-m^2}\right)^2
                          \bigl(\frac{m^2}{t}\bigr)^2
                                          \Biggr] \ .
\end{eqnarray}
As in the massless case, the cross section for $\bar q +g \to \bar q + G_{KK}$ is also
 the same as that of $q +g \to q + G_{KK}$. 
  For the process
 $g+g \to g+ G_{KK}$, we find 
 \begin{eqnarray}
\frac{d\sigma}{d \cos\theta}(g+g \to g+ G_{KK})&=& \frac{3}{16}
      \frac{\pi \alpha_s G_N}{(1-m^2/s)(1-\cos^2\theta)}\Biggl[(3 + 
   \cos^2\theta )^2\left(1 + \bigl(\frac{m^2}{s}\bigr)^4\right) \\ \nonumber
     && - 4 ( 7 + \cos^4\theta)\frac{m^2}{s }\left(1 +  
           \bigl(\frac{m^2}{s}\bigr)^2\right) 
          + 6 (9 - 2  \cos^2\theta +  \cos^4\theta)  
                      \bigl(\frac{m^2}{s}\bigr)^2  
                                          \Biggr] \ .
\end{eqnarray}
The first observation is that the angular dependence is different in the massless graviton and in the Kaluza Klein graviton case. By studying the angular distribution of the jets, one could measure the mass of the graviton carrying the missing energy and thus differentiate the four dimensional model from the extra-dimensional model. We shall now study the limit $m\to 0$ of the cross-sections for the production of massive Kaluza-Klein modes. One finds:
\begin{eqnarray}
 \frac{d\sigma}{d \cos\theta}(q + \bar q \to g + G_{KK})&=& \frac{1}{72 \pi}
 \frac{g_s^2 }{\bar M^2} \left ( \frac{s^2 +2t^2+ 2 t s}{s^2} \right ) \ \mbox{for $m \to 0$},
 \end{eqnarray}
for $q + \bar q \to g + G_{KK}$
 \begin{eqnarray}
\frac{d \sigma}{d \cos \theta}(q +  g \to q + G_{KK}) = - \frac{g_s^2 (2 s^2+ 2 s t + t^2)}{ 192 \pi \bar M^2 s t }  \ \mbox{for $m \to 0$},
\end{eqnarray}
for $q +  g \to q + G_{KK}$ and finally
 \begin{eqnarray}
\frac{d \sigma}{d \cos \theta}(g +  g \to g + G_{KK}) = - \frac{3 g_s^2 (s^2+s t + t^2)^2}{ 128 \pi \bar M^2 s^2 t (s+t)}  \ \mbox{for $m \to 0$},
\end{eqnarray}
for $g +  g \to g + G_{KK}$.   The smoothness of this limit has been pointed out in \cite{Mirabelli:1998rt} where it is explained that because of helicity conservation, only the polarizations modes that correspond to the massless graviton are produced in that reaction.

\section{Conclusions}

We have considered the production at the LHC of gravitons in a four dimensional model with a scale of quantum gravity in the TeV region. The graviton which appears in that model is the usual graviton of general relativity and it is massless. Because of the renormalization group evolution of Newton's constant, the coupling of the graviton to Standard Model matter becomes strong in collisions of particles in the TeV regime. We compare our calculations to those that have been done in the framework of extra dimensional scenarios. In both massive and massless graviton case, the signature is jets plus missing energy. However, the angular distribution of the jets would allow to differentiate between our four dimensional model and models with large extra dimensions. The next natural extension of the work would  be to implement these results in an event generator and to study the background. This is work in progress.

\bigskip

\section*{Acknowledgments} 
 This work  is supported in part by the Belgian Federal Office for Scientific, Technical and Cultural Affairs through the Interuniversity Attraction Pole P6/11. 

\section*{Appendix A: On the renormalization of the Planck Mass}

In this appendix, we shall justify at the hand waving level that the extrapolation $\mu \to \mu_*$ is reliable. The first question is that of the graviton's contribution to the renormalization of the Planck mass.
It is possible to calculate the contribution of the graviton to the running of the Planck mass using the effective theory of  general relativity  developed by Donoghue \cite{Donoghue:1994dn}. This beautiful and difficult calculation done by Donoghue and collaborators  \cite{BjerrumBohr:2002ks} gives:
 \begin{eqnarray}
 \label{Nrunninggrav}
M(\mu)^2=M(0)^2 -\frac{\mu^2}{12 \pi} \left (-\frac{334}{5} \right)
\end{eqnarray}
where we have identified $\mu = r^{-1}$ in which $r$ is the distance from the source of the potential. There is obviously some numerical uncertainty in the value of the cutoff, but the important result is that quantum gravitational interactions make the Planck mass bigger at high energy.

The full renormalized Planck mass at one loop is thus
\begin{eqnarray}
M(\mu)^2=M(0)^2 -\frac{\mu^2}{12 \pi} \left (N_0+N_{1/2}-4N_1-\frac{334}{5} \right).
\end{eqnarray}
If the number of matter fields $N$ is large, clearly quantum gravitational effects are a $1/N$ correction:
\begin{eqnarray}
M(\mu)^2=M(0)^2 -\frac{\mu^2}{12 \pi} N \left (1-\frac{334}{5} \frac{1}{N} \right)
\end{eqnarray}
and thus under control since they represent a small correction to the one loop result. This limit has been studied already by Tomboulis \cite{Tomboulis:1977jk} and later by Smolin \cite{Smolin:1981rm}.

\begin{figure}
$$
\includegraphics[width=7cm]{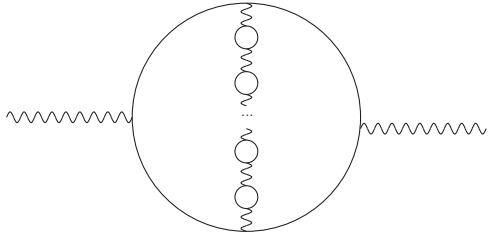}$$
\caption{\label{FG1}\em
Higher loops contribution to the renormalization of the Planck mass. The wavy lines represent gravitons, whereas continuous lines are matter field loops. For a given number of matter loops, the most important contributions comes from the diagram where a single graviton propagator contains all the matter field loops}
\end{figure}

\begin{figure}
$$
\includegraphics[width=7cm]{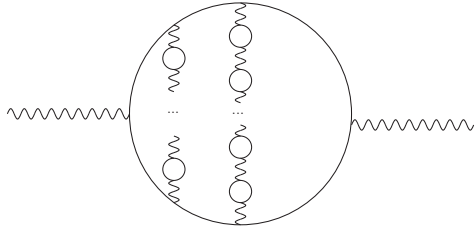}$$
\caption{\label{FG2}\em
Higher loops contribution to the renormalization of the Planck mass. For a fixed number of loops this topology is suppressed compared to the diagram shown in Figure (\ref{FG1}).}
\end{figure}

One may wonder about higher order loop corrections such as diagrams depicted in Figures (\ref{FG1}) and (\ref{FG2}). A back of the envelop calculation shows that diagrams of the type depicted in Figure (\ref{FG1}) lead to a contribution of the type:
\begin{eqnarray}
\sim \frac{1}{(4 \pi^2)^{l+2}}  N^l \left (\frac{\Lambda}{M(0)} \right)^{2l}  \frac{\Lambda^4 N}{M(0)^2}
\end{eqnarray}
where $l$ is the number of matter fields loop on the graviton line and $\Lambda$ is a dimensionful cutoff. These contributions are  small compared to the first loop result ($M(0) \sim 10^{18}$ GeV and $\Lambda \sim 10^3$ GeV). In other words if we were able to resum the diagrams on the graviton line, we would obtain a graviton line with a coupling to matter only suppressed $M(\mu_*)$ but not enhanced by $N$.

Diagrams with more graviton propagators (see e.g. Figure (\ref{FG2})), for a given number of matter field loops, are suppressed compared to those shown in Figure (\ref{FG1}):
\begin{eqnarray}
 \sim \frac{1}{(4 \pi^2)^{l+3}}  N^l \left (\frac{\Lambda}{M(0)} \right)^{2l}  \frac{\Lambda^6 N}{M(0)^4}.
\end{eqnarray}


These results can be applied to the large hidden sector model considered in this paper. Lowering the scale of strong gravity $\mu_\star$ to the TeV region requires introducing a large hidden sector with $10^{33}$ new particles of spin 0 and/or 1/2 that only interact gravitationally with the Standard Model fields. Clearly with such a large hidden sector we are in the limit where loops of gravitons are irrelevant and higher loops involving matter fields are under control, as discussed above.

This result also has implications for grand unified theories. It has be shown \cite{Calmet:2008df} that in typical supersymmetric theory of grand unified based on for example SO(10), the large number of particles with masses close to the unification scale ($N\sim 1000$) leads to a shift of the strength at which gravity becomes strong. One finds $\mu_\star \sim 10^{17}$ GeV rather than  $\mu_\star \sim 10^{18}$ GeV which implies that operators induced by strong gravitational effects can dramatically impact the unification conditions of the gauge  couplings of the Standard Model.

\section*{Appendix B: Feynman rules}

In this appendix we summarize the Feynman rules we have used in our calculations. We used the conventions of \cite{Han:1998sg}:
\begin{align}
&C_{\mu\nu,\rho\sigma} & =\quad & \eta_{\mu\rho} \eta_{\nu\sigma} + \eta_{\mu\sigma} \eta_{\nu\rho} -\eta_{\mu\nu} \eta_{\rho\sigma}\\ 
\nonumber \\ 
&D_{\mu\nu,\rho\sigma}(k_1,k_2) & =\quad & \eta_{\mu\nu} k_{1\sigma} k_{2\rho} - 
\eta_{\mu\sigma} k_{1\nu} k_{2\rho}- \eta_{\mu\rho} k_{1\sigma} k_{2\nu} + \eta_{\rho\sigma} k_{1\mu} k_{2\nu}   \\ \nonumber &&&   - 
\eta_{\nu\sigma} k_{1\mu} k_{2\rho}- \eta_{\nu\rho} k_{1\sigma} k_{2\mu} + \eta_{\rho\sigma} k_{1\nu} k_{2\mu}  \\ \nonumber
\\ 
&E_{\mu\nu,\rho\sigma}(k_1,k_2) & =\quad & \eta_{\mu\nu} ( k_{1\rho} k_{1\sigma} + k_{2\rho} k_{2\sigma}+k_{1\rho} k_{2\sigma}) 
\\ \nonumber &&&   
-\eta_{\nu\sigma} k_{1\mu}k_{1\rho} -\eta_{\nu\rho} k_{2\mu}k_{2\sigma} 
 -\eta_{\mu\sigma} k_{1\nu}k_{1\rho} -\eta_{\mu\rho} k_{2\nu}k_{2\sigma} 
\end{align}
The propagator for the quarks and gluons are respectively given by
\begin{eqnarray}
\frac{i(\fmslash p +m)}{p^2-m^2+i \epsilon}
\end{eqnarray}
and
\begin{eqnarray}
\frac{-i \delta^{ab}}{k^2} \left (g^{\mu\nu}- \frac{k^\mu k^\nu}{k^2} (1-\xi)\right).
\end{eqnarray}

The vertices describing the interactions of the graviton are given by

\begin{tabular}{c>{\raggedright}m{5in}}
 & 
 \tabularnewline
\includegraphics[scale=0.25]{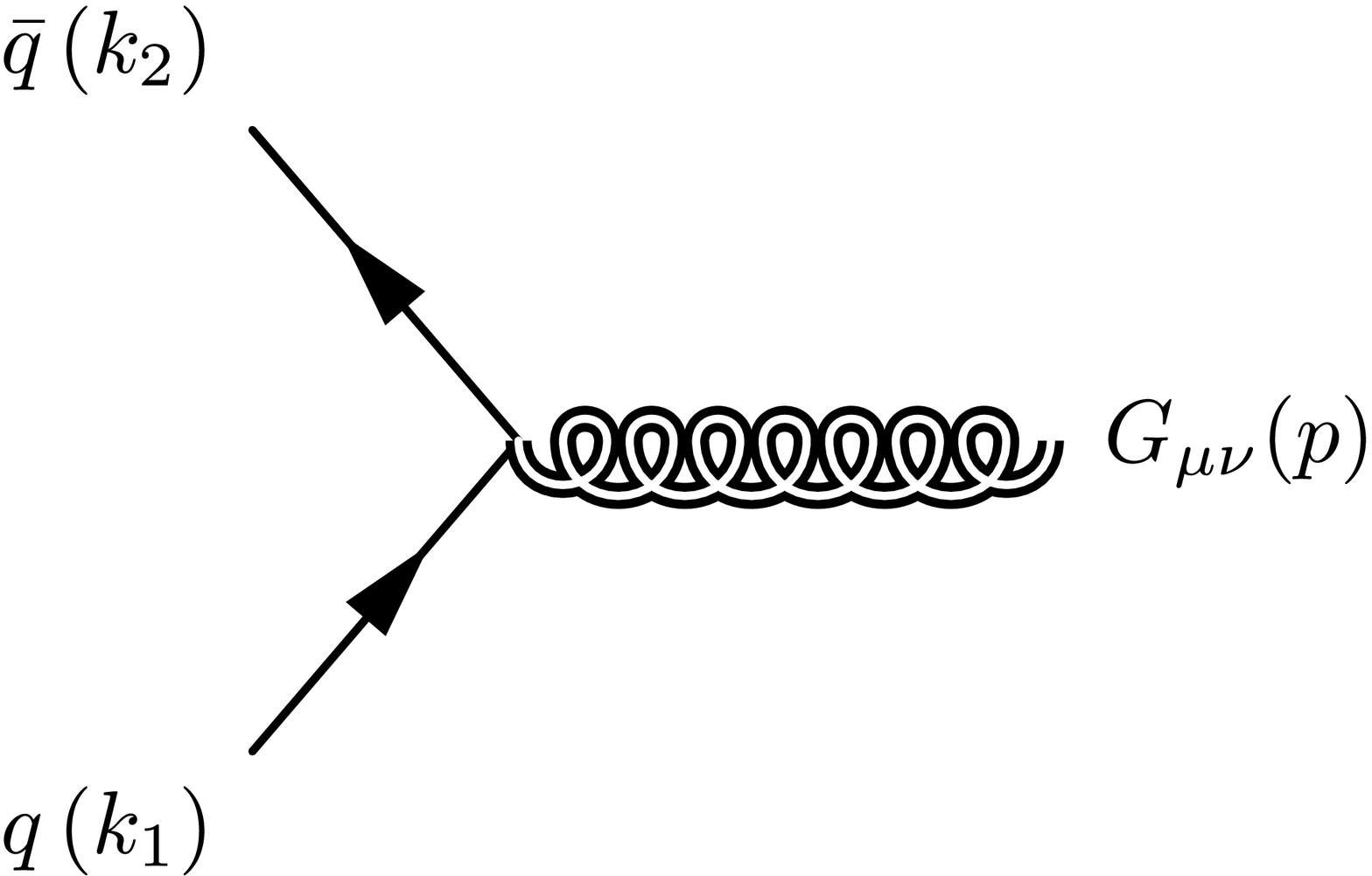} & $=-i\,\frac{\kappa}{8}\left\{ \gamma_{\mu}\left(k_{1\,\nu}+k_{2\,\nu}\right)+\gamma_{\nu}\left(k_{1\,\mu}+k_{2\,\mu}\right)-2\eta_{\mu\nu}\left(\not k_{1}+\not k_{2}\right)\right\} $
$\qquad$$\qquad$\vspace{1in}
\tabularnewline
 & \tabularnewline
\includegraphics[scale=0.25]{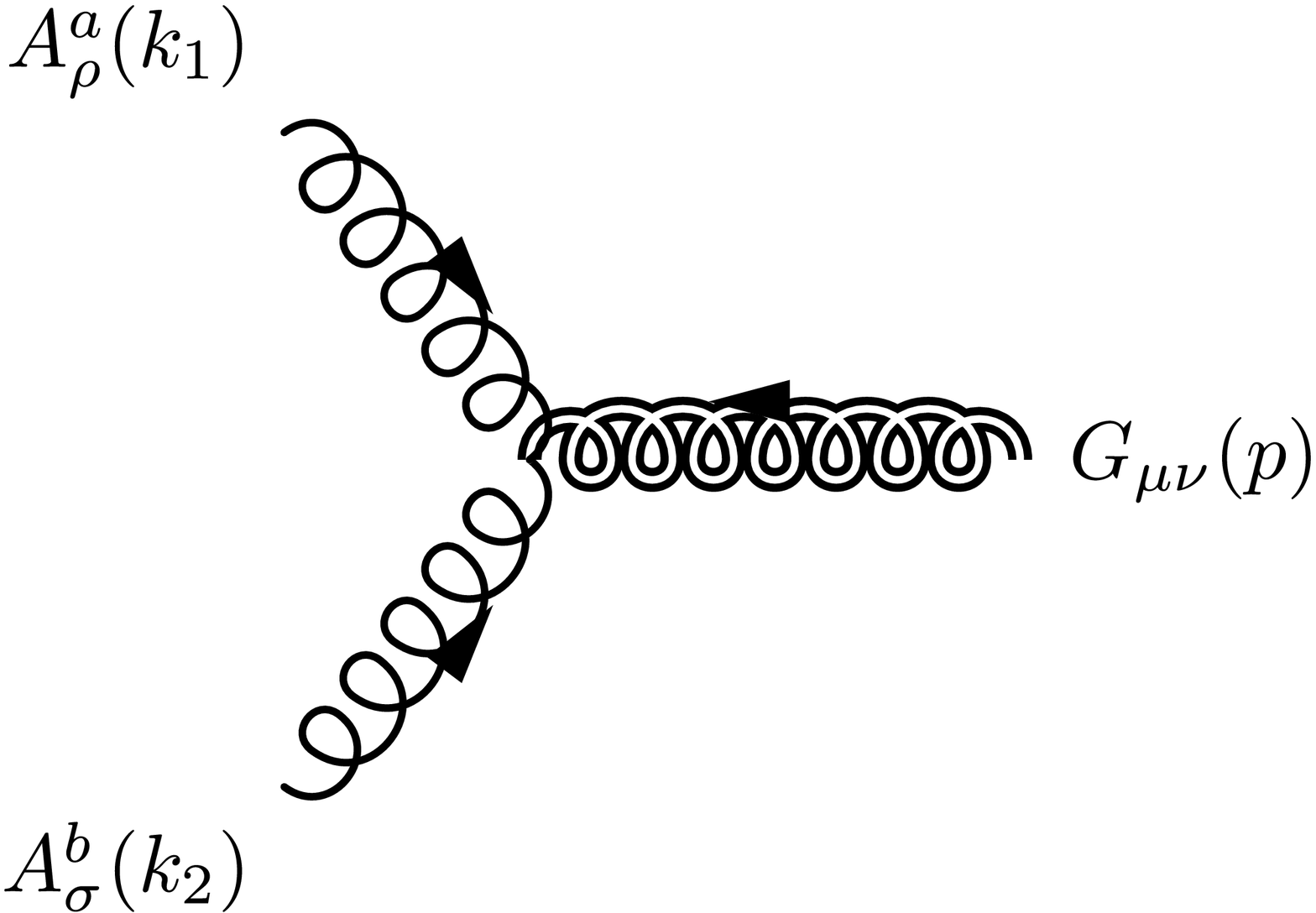} & $=i\,\frac{\kappa}{2}\delta^{ab}\left\{ \left[k_{1}\cdot k_{2}\right]C_{\mu\nu\rho\sigma}+D_{\mu\nu\rho\sigma}\left(k_{1},k_{2}\right)+\xi^{-1}E_{\mu\nu\rho\sigma}\left(k_{1},k_{2}\right)\right\} $
$\qquad$$\qquad$\vspace{1in}
\tabularnewline
& \tabularnewline
\includegraphics[scale=0.25]{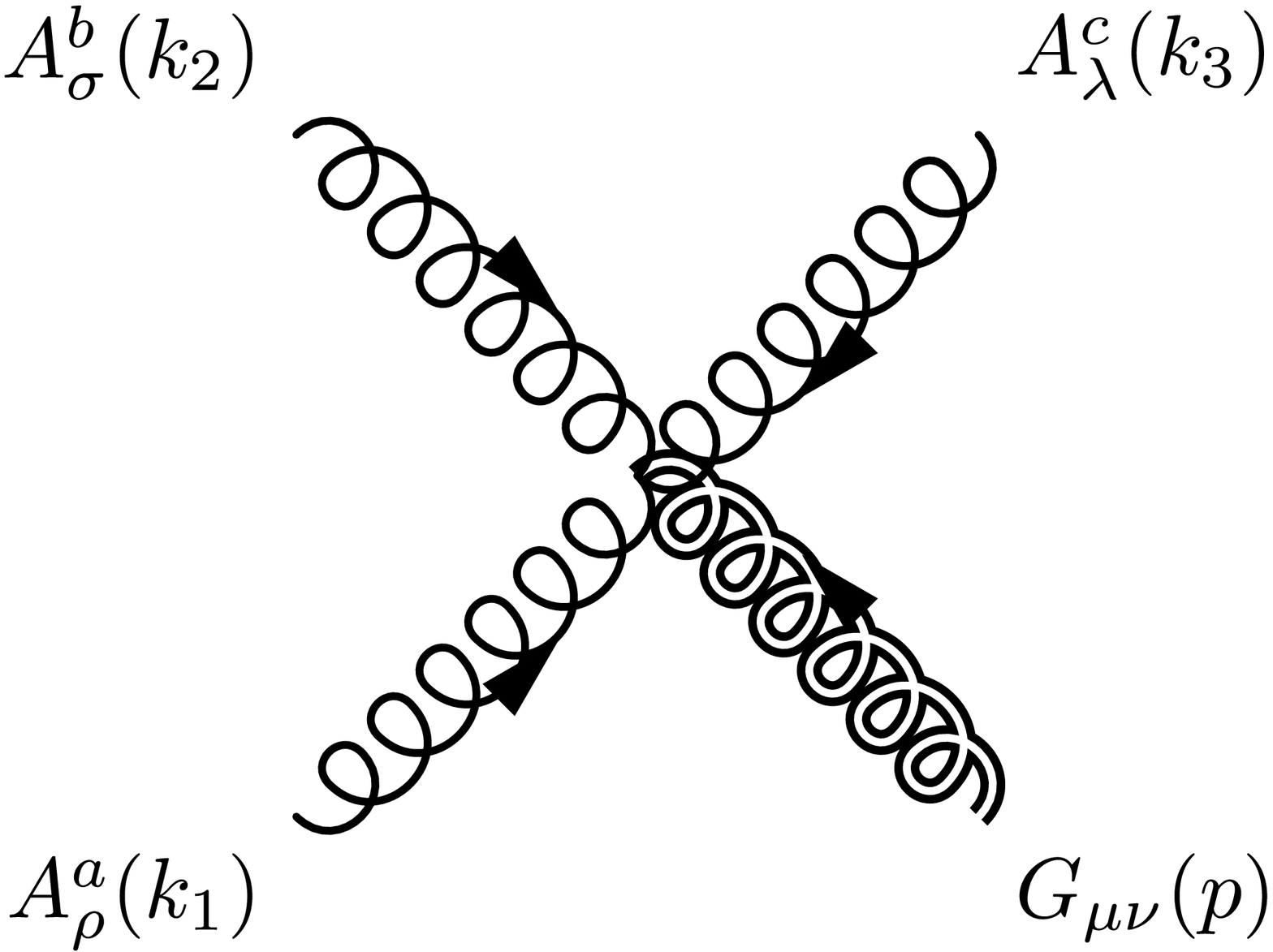} & $=g\,\frac{\kappa}{2}f^{abc}\left\{ C_{\mu\nu\rho\sigma}\left[k_{1\,\lambda}-k_{2\,\lambda}\right]+C_{\mu\nu\rho\lambda}\left[k_{3\,\sigma}-k_{1\,\sigma}\right]\right.$

\medskip{}

$\qquad\qquad\left.+C_{\mu\nu\sigma\lambda}\left[k_{2\,\rho}-k_{3\,\rho}\right]+F_{\mu\nu\rho\sigma\lambda}\left(k_{1},\, k_{2},\, k_{3})\right)\right\} $
$\qquad$$\qquad$\vspace{1in}
\tabularnewline
\end{tabular}

\begin{tabular}{c>{\raggedright}m{5in}}
 \tabularnewline
\includegraphics[scale=0.25]{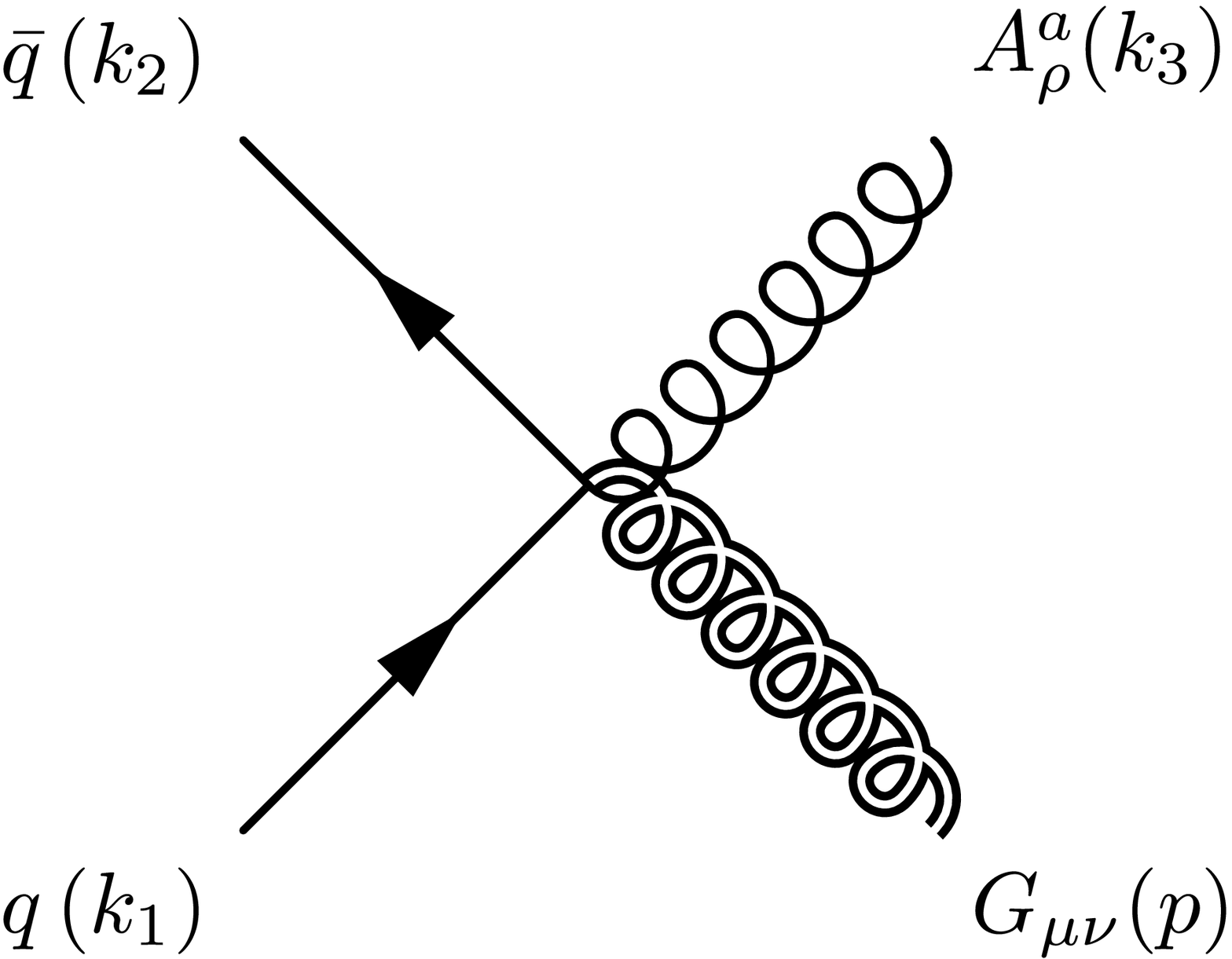} & $=i\, g\,\frac{\kappa}{4}\, T^{a}\left\{ C_{\mu\nu\rho\sigma}-\eta_{\mu\nu}\eta_{\rho\sigma}\right\} \gamma^{\sigma}$
$\qquad$$\qquad$\vspace{1in}
\tabularnewline
 & \tabularnewline
\end{tabular}
with the understanding that $\kappa= 16 \pi G_N$ is scale dependent. $\xi$ is a gauge fixing parameter.
%\pagebreak{}

Finally, we also make use of the standard three particles vertices:

\begin{center}
\textbf{\textsc{\LARGE }}\begin{tabular}{c>{\raggedright}m{5in}}
 & \tabularnewline
\includegraphics[scale=0.25]{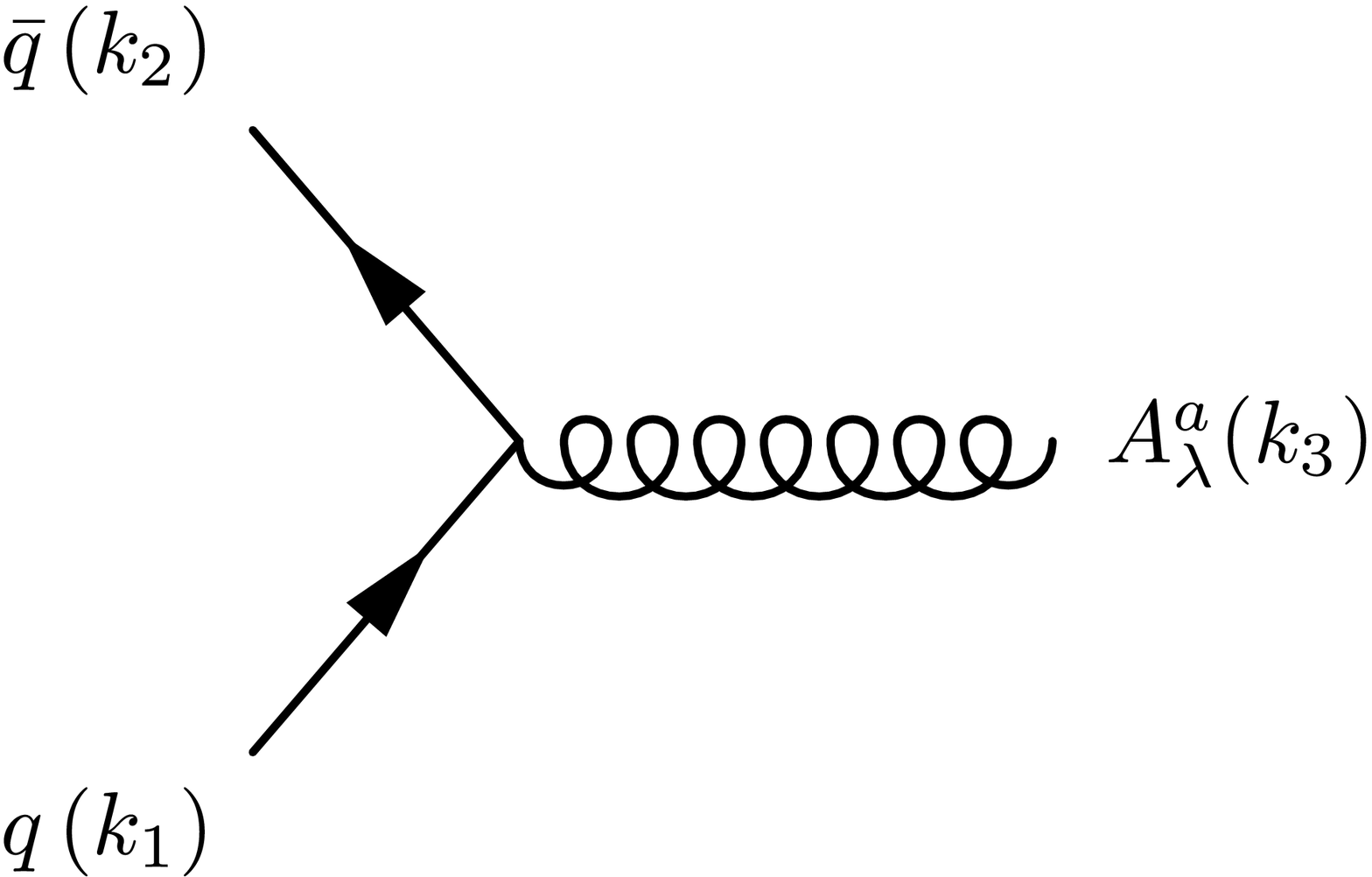} & $=i\, g\, T^{a}\,\gamma^{\lambda}$ $\qquad$$\qquad$$\qquad$\vspace{1in}
\tabularnewline
 & $ $\tabularnewline
\includegraphics[scale=0.25]{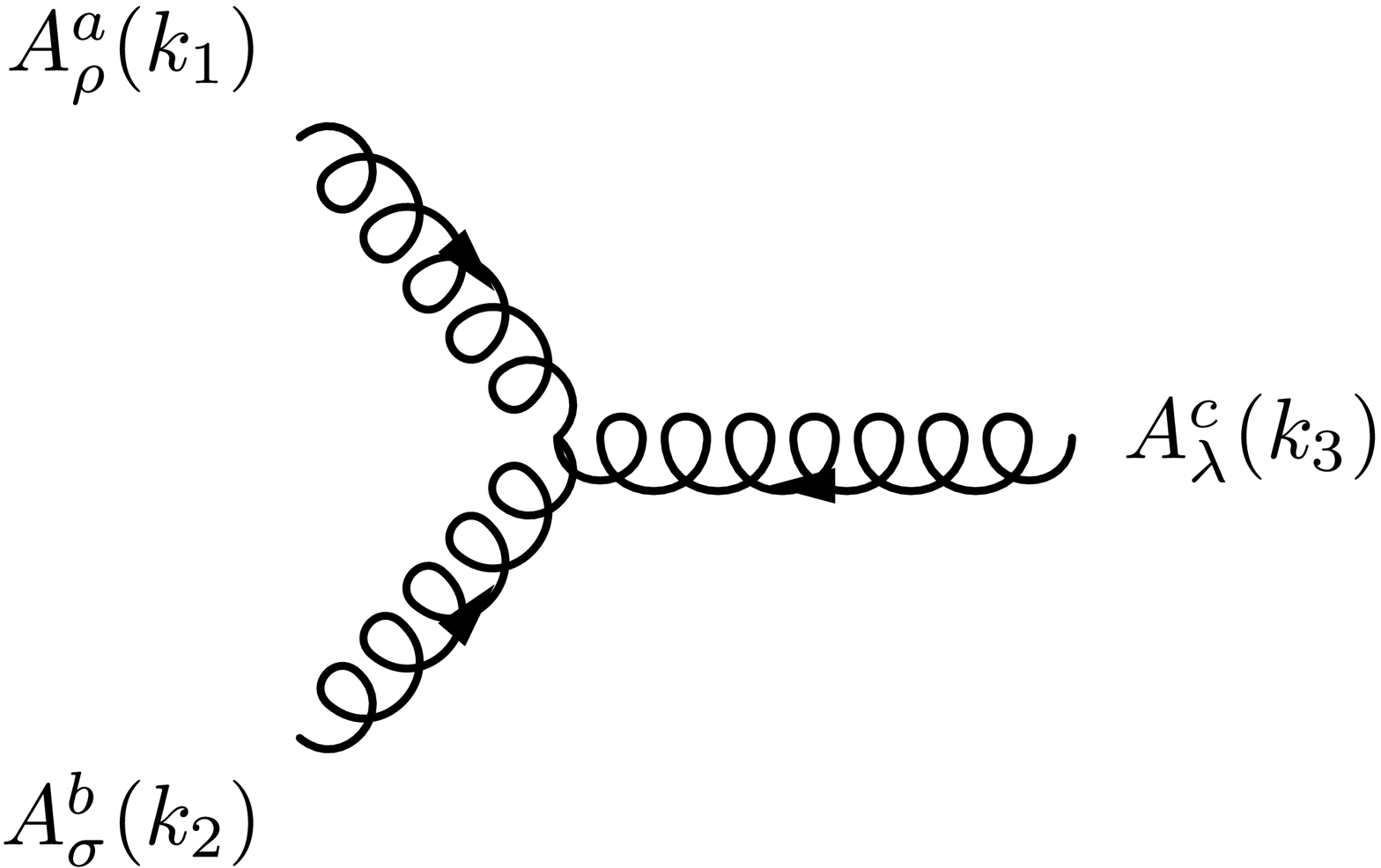} & $=g\, f^{abc}\left\{ \eta_{\rho\sigma}\left(k_{1\,\lambda}-k_{2\,\lambda}\right)+\eta_{\sigma\lambda}\left(k_{2\,\rho}-k_{3\,\rho}\right)+\eta_{\lambda\rho}\left(k_{3\,\sigma}-k_{1\,\sigma}\right)\right\} $
$\qquad$$\qquad$\vspace{1in}
\tabularnewline
 & \tabularnewline
\end{tabular}
\par\end{center}

\bigskip
\bigskip
%%%%%%%% End comments

%\newpage

%%%%%%%%%%%%%%%%%%%%%%%%%%%%%%%%%%%%%%%%%%%%%%%%%%%%%%%%%%%%%%%%%
%%%
%%%                     BIBLIOGRAPHY
%%%
%%%%%%%%%%%%%%%%%%%%%%%%%%%%%%%%%%%%%%%%%%%%%%%%%%%%%%%%%%%%%%%%%

\bigskip

%\newpage
%\vskip .75 in

\end{document}